\documentclass[iop,apj,numberedappendix,twocolappendix]{emulateapj}

\usepackage{apjfonts}
\usepackage{natbib}
\usepackage{amssymb}
\usepackage{enumitem}
\usepackage[breaklinks,colorlinks,citecolor=blue,linkcolor=magenta]{hyperref}

\newcommand{\magiicat}{\hbox{{\rm MAG}{\sc ii}CAT}}
\newcommand{\MgIIdblt}{{\rm Mg}\kern 0.1em{\sc ii}~$\lambda\lambda 2796, 2803$}
\newcommand{\MgII}{\hbox{{\rm Mg}\kern 0.1em{\sc ii}}}
\newcommand{\OVIdblt}{{\rm O}\kern 0.1em{\sc vi}~$\lambda\lambda 1031, 1037$} 
\newcommand{\OVI}{\hbox{{\rm O}\kern 0.1em{\sc vi}}}
\newcommand{\OVII}{\hbox{{\rm O}\kern 0.1em{\sc vii}}}
\newcommand{\OVIII}{\hbox{{\rm O}\kern 0.1em{\sc viii}}}
\newcommand{\NeVIII}{\hbox{{\rm Ne}\kern 0.1em{\sc viii}}}
\newcommand{\CIV}{\hbox{{\rm C}\kern 0.1em{\sc iv}}}
\newcommand{\HI}{\hbox{{\rm H}\kern 0.1em{\sc i}}}
\newcommand{\kms}{\hbox{km~s$^{-1}$}}
\newcommand{\etal}{et~al.}

\newcommand{\vfifty}{\hbox{$\Delta v(50)$}}
\newcommand{\vninety}{\hbox{$\Delta v(90)$}}
\newcommand{\blue}{\hbox{$B-K<1.66$}}
\newcommand{\red}{\hbox{$B-K\geq1.66$}}

\newcommand{\face}{\hbox{$i < 51^{\circ}$}}
\newcommand{\edge}{\hbox{$i \geq 51^{\circ}$}}
\newcommand{\major}{\hbox{$\Phi < 45^{\circ}$}}
\newcommand{\minor}{\hbox{$\Phi \geq 45^{\circ}$}}

\shorttitle{OVI CGM Kinematics}
\shortauthors{\sc Nielsen {\etal}}
\slugcomment{Accepted for publication in ApJ on November 22nd, 2016}

\begin{document}

\title{The Highly Ionized Circumgalactic Medium is Kinematically
  Uniform Around Galaxies}

\author{
Nikole M. Nielsen$^{1}$,
Glenn G. Kacprzak$^{1}$,
Sowgat Muzahid$^{2,3}$,
Christopher W. Churchill$^{4}$,
Michael T. Murphy$^1$,
and
Jane C. Charlton$^2$
}

\affil{$^1$ Centre for Astrophysics and Supercomputing, Swinburne
  University of Technology, Hawthorn, Victoria 3122, Australia;
  nikolenielsen@swin.edu.au\\
  $^2$ Department of Astronomy \& Astrophysics, The Pennsylvania State
  University, State College, PA 16801, USA\\
  $^3$ Leiden Observatory, Leiden University, PO Box 9513, NL-2300 RA
  Leiden, The Netherlands\\
  $^4$ Department of Astronomy, New Mexico State University, Las
  Cruces, NM 88003, USA}

\begin{abstract}

The circumgalactic medium (CGM) traced by {\OVIdblt} doublet
absorption has been found to concentrate along the projected major and
minor axes of the host galaxies. This suggests that {\OVI} traces
accreting and outflowing gas, respectively, which are key components
of the baryon cycle of galaxies. We investigate this further by
examining the kinematics of 29 {\OVI} absorbers associated with
galaxies at $0.13 < z_{\rm gal} < 0.66$ as a function of galaxy color,
inclination, and azimuthal angle. Each galaxy was imaged with {\it
  HST} and the absorption was detected in COS/{\it HST} spectra of
nearby ($D<200$~kpc) background quasars. We use the pixel-velocity
two-point correlation function to characterize the velocity spread of
the absorbers, which is a method used previously for a sample of
{\MgII} absorber--galaxy pairs. The absorption velocity spread for
{\OVI} is more extended than {\MgII}, which suggests that the two ions
trace differing components of the CGM. Also contrary to {\MgII}, the
{\OVI} absorption velocity spreads are similar regardless of galaxy
color, inclination, and azimuthal angle. This indicates that the
kinematics of the high ionization gas is not strongly influenced by
the present star formation activity in the galaxy. The kinematic
homogeneity of {\OVI} absorption and its tendency to be observed
mainly along the projected galaxy major and minor axes is likely due
to varying ionization conditions and gas densities about the
galaxy. Gas in intermediate azimuthal angles may be ionized out of the
{\OVI} phase, possibly resulting in an azimuthal angle dependence on
the distribution of gas in higher ionization states.

\end{abstract}


\section{Introduction}
\label{sec:intro}

The circumgalactic medium (CGM) is a massive reservoir of multiphase
gas surrounding a galaxy, with a gas mass comparable to the gas mass
in the galaxy itself \citep{thom11, tumlinson11, werk13,
  peeples14}. It regulates the star formation rate of the galaxy
through a balance of inflows and outflows into, out of, and through
the CGM \citep[e.g.,][]{oppenheimer08, lilly-bathtub}. For these
reasons, understanding the multiphase nature, locations, and kinematic
properties of gas in the CGM is crucial to understanding how galaxies
evolve to form the galaxies observed today. 

Much of the current understanding of the CGM comes from the
low-ionization {\MgIIdblt} doublet absorption in background quasar
spectra due to it being easily observable from the ground in optical
wavelengths at $z\sim 1$. Recent work has found that {\MgII} absorbers
are preferentially located along the projected major and minor axes of
their host galaxies \citep{bordoloi11, bouche12, kcn12, lan14} and
their kinematics show distinct differences with galaxy orientation,
color, and other properties \citep{magiicat5, magiicat4}. {\MgII} is
commonly associated with outflows \citep[e.g.,][]{rubin-winds,
  rubin-winds14, bouche12, martin12, bordoloi14-model, bordoloi14,
  kacprzak14} and accretion or recycled outflows
\citep[e.g.,][]{steidel02, ggk-sims, stewart11, martin12,
  rubin-accretion, bouche13, ford14}.

However, given the multiphase nature of the CGM, {\MgII} traces only a
fraction of the CGM. The high-ionization {\OVIdblt} doublet absorption
is another common tracer of the CGM. \citet{tumlinson11} has shown
that the presence of {\OVI} is governed by the star formation rate of
the host galaxy, with more absorbers associated with star forming
galaxies and more nonabsorbers with passive galaxies. {\OVI} has been
further studied extensively \citep[e.g.,][]{wakker09, prochaska11,
  johnson13, johnson15, stocke13, mathes14, savage14}, but the
physical processes giving rise to the gas traced by {\OVI} is still
debated.

Recently, \citet{oppenheimer16} examined circumgalactic oxygen in the
EAGLE simulations. They found that {\OVI} is not the dominant
ionization state of oxygen in galaxy halos, and that the column
densities of {\OVI} peak for $L_{\ast}$ galaxies, with lower column
densities for lower and higher mass halos. Given this, the authors
suggest that {\OVI} is primarily a tracer of the virial temperature of
a galaxy, where $L_{\ast}$ galaxies have a virial temperature that
results in the largest {\OVI} ionization fraction. For galaxies less
massive than an $L_{\ast}$ galaxy, the virial temperature is too cool
for strong {\OVI} and more massive galaxies ionize the oxygen into
higher ionization states. The authors also found no connection between
star formation and the {\OVI} out to 150~kpc, where the median ``age''
of {\OVI} is greater than 5 Gyrs. Consequently, the
\citet{tumlinson11} results may be reflecting the changing ionization
conditions with galaxy mass rather than a star formation rate
dependence.

Comparing the properties of and processes depositing both the
low-ionization gas traced by {\MgII} and the high-ionization gas
traced by {\OVI} have become more common. For example,
\citet{muzahid15} studied an absorber--galaxy pair in detail, where
the pair has both {\MgII} and {\OVI} absorption probed along the minor
axis of an edge-on galaxy. The authors concluded that the low- and
high-ionization absorption traced recycled accretion and outflows,
respectively. The metallicities of the ions are different, with the
high-ionization phase having a metallicity (super-solar) over an order
of magnitude greater than the low-ionization phase.

Using mock quasar absorption-line observations in hydrodynamic
cosmological simulations, \citet{churchill15} examined the properties
of the multiphase gas in the circumgalactic medium of a dwarf
galaxy. The authors traced the line-of-sight spatial locations of the
cells that dominate the absorption profiles. They found that while
{\CIV} and {\OVI} are observed at similar velocities as {\HI} and
{\MgII}, the higher ionization gas traces different structures in
different locations (spread over up to 100~kpc) along the line of
sight. Also studying the simulated CGM, \citet{ford14} found that
{\OVI} primarily traces ``ancient outflows'' in which the gas was
ejected from the galaxy by outflows greater than 1 Gyr prior to
$z=0.25$. In contrast, they found that {\MgII} is dominated by
recycled accretion.

\citet{kacprzak15} measured the orientations (inclinations and
azimuthal angles) of galaxies associated with both {\OVI} absorbers
and non-absorbers. They define an azimuthal angle of $\Phi=0^{\circ}$
as having the background quasar sightline aligned with the projected
galaxy major axis, and $\Phi=90^{\circ}$ as the sightline along the
projected galaxy minor axis. The authors reported that detected {\OVI}
absorption is preferentially found along the major and minor axes of
the host galaxy \citep[similar to the behavior of {\MgII};][]{kcn12},
suggesting that the absorbers in these regions traced
accretion/recycling and outflows, respectively. Absorption was rarely
detected within azimuthal angles of $30^{\circ}$ to $60^{\circ}$, with
the authors proposing that {\OVI} is not mixed throughout the CGM. The
equivalent widths of absorption were also found to be greater along
the minor axis than the major axis, hinting that the velocity spreads,
the column densities, or both were dependent on the azimuthal angle at
which gas is probed. To further examine the physics involved and the
gas properties in relation to the galaxy, we study the kinematics of
these {\OVI} absorbers here.

Both \citet{magiicat4} and \citet{magiicat5} (hereafter {\magiicat} IV
and {\magiicat} V, respectively) used the pixel-velocity two-point
correlation function (TPCF) method for {\MgII} absorbers to examine
the velocity spreads as a function of galaxy color, redshift, impact
parameter, inclination, and azimuthal angle. The TPCF method produced
clear results in which the greatest absorber velocity dispersions were
located along the projected minor axis ($\Phi\geq 45^{\circ}$) of
face-on ($i<57^{\circ}$) blue galaxies. These large velocity
dispersions were attributed to bipolar outflows, which, for the
largest velocity dispersions, are pointed nearly directly towards or
away from the observer. Red galaxies did not show these large
dispersions, and in fact, showed the smallest velocity dispersions out
of all subsamples, indicating a lack of outflowing material in
{\MgII}. In this paper, we now use this TPCF method on a sample of 29
{\OVI} absorbing galaxies as a function of galaxy color, inclination,
and azimuthal angle. We also compare the {\OVI} TPCFs to the previous
results with {\MgII}.

This paper is organized as follows: Section~\ref{sec:methods}
describes the {\OVI} sample and data analysis to obtain the galaxy and
absorption properties. We also briefly describe the pixel-velocity
two-point correlation function (TPCF)
method. Section~\ref{sec:results} presents the results of examining
the TPCFs of {\OVI} absorbers as a function of galaxy color, azimuthal
angle, and inclination. We also present the TPCFs for our {\MgII}
sample ({\magiicat} IV and V) in this section for comparison between
ions. In Section~\ref{sec:discussion} we discuss our results in the
context of previous work. Finally, Section~\ref{sec:conclusions}
summarizes and concludes our findings. Throughout the paper we use AB
magnitudes and a $\Lambda$CDM cosmology ($H_0=70$~km s$^{-1}$
Mpc$^{-1}$, $\Omega_M=0.3$, $\Omega_{\Lambda}=0.7$).

\section{Sample and Data Analysis}
\label{sec:methods}

In this section, we describe the galaxy properties and the quasar
spectra for our {\OVI} absorber--galaxy sample, which is the focus of
this paper. We also briefly describe the {\MgII} sample published in
{\magiicat} IV and V, which we use here for comparison to the more
highly ionized {\OVI} CGM. Finally, we briefly describe our
pixel-velocity TPCF method for studying the absorption kinematics.

\subsection{Galaxy Properties}
\label{sec:galprops}

We use the subset of 29 galaxies with colors and detected {\OVIdblt}
doublet absorption from the sample presented in \citet{kacprzak15}
(hereafter Kacprzak15), which were identified as part of our
``Multiphase Galaxy Halos'' large {\it HST} program \citep[e.g.,
  Kacprzak15,][]{muzahid15, muzahid16} or obtained from the
literature. The galaxies with non-detections in {\OVI} reported in
Kacprzak15 are not included here as we cannot measure their absorption
kinematics. The absorption-selected galaxies are located at
spectroscopic redshifts of $0.13 < z_{\rm gal} < 0.66$ (median
$\langle z_{\rm gal} \rangle = 0.244$) and within an on-the-sky
projected distance of $D \sim 200$~kpc (21.1~kpc$<D<203.2$~kpc,
$\langle D \rangle =93.2$~kpc) from a background quasar. These are
isolated galaxies, where no neighbors were identified within a
projected distance of 200~kpc from the quasar line-of-sight and within
a line-of-sight velocity separation of 500~{\kms}.

%%%%%%%%%%%%%%%%%%%%%
\begin{deluxetable}{llrccc}
 \tablecolumns{4}
 \tablewidth{0pt}
 \tablecaption{Galaxy Properties \label{tab:galprops}}
 \tablehead{
   \colhead{Field} &
   \colhead{$z_{\rm gal}$} &
   \colhead{$D$} &
   \colhead{$B-K$} &
   \colhead{$i$} &
   \colhead{$\Phi$} \\
   \colhead{} &
   \colhead{} &
   \colhead{(kpc)} &
   \colhead{} &
   \colhead{(deg)} &
   \colhead{(deg)}
   }
 \startdata

J012528$-$000555 & 0.3985 & 163.0 & 1.80 & 63.2 & 59.3 \\[3pt]
J035128$-$142908 & 0.3567 &  72.2 & 0.28 & 28.5 &  4.8 \\[3pt]
J045608$-$215909 & 0.3818 & 103.4 & 1.78 & 57.1 & 63.7 \\[3pt]
J045608$-$215909 & 0.4847 & 108.0 & 1.66 & 42.1 & 85.2 \\[3pt]
J091440$+$282330 & 0.2443 & 105.8 & 1.48 & 38.9 & 18.2 \\[3pt]
J094331$+$053131 & 0.3530 &  96.4 & 1.40 & 44.3 &  8.1 \\[3pt]
J094331$+$053131 & 0.5480 & 150.8 & 1.17 & 58.8 & 67.1 \\[3pt]
J095000$+$483129 & 0.2119 &  93.5 & 3.13 & 47.7 & 16.6 \\[3pt]
J100402$+$285535 & 0.1380 &  56.7 & 1.21 & 79.1 & 12.3 \\[3pt]
J100902$+$071343 & 0.2278 &  63.9 & 1.39 & 66.2 & 89.5 \\[3pt]
J104116$+$061016 & 0.4432 &  56.2 & 2.81 & 49.8 &  4.2 \\[3pt]
J111908$+$211918 & 0.1380 & 137.9 & 2.21 & 26.3 & 34.4 \\[3pt]
J113327$+$032719 & 0.1545 &  55.6 & 1.53 & 23.5 & 56.0 \\[3pt]
J113910$-$135043 & 0.2044 &  93.1 & 2.30 & 83.4 &  5.8 \\[3pt]
J113910$-$135043 & 0.2123 & 174.8 & 2.10 & 84.9 & 80.4 \\[3pt]
J113910$-$135043 & 0.2198 & 121.9 & 2.42 & 85.0 & 44.9 \\[3pt]
J113910$-$135043 & 0.3191 &  73.2 & 1.60 & 83.3 & 39.0 \\[3pt]
J123304$-$003134 & 0.3185 &  88.9 & 1.63 & 38.6 & 17.0 \\[3pt]
J124154$+$572107 & 0.2053 &  21.1 & 1.67 & 56.4 & 77.6 \\[3pt]
J124154$+$572107 & 0.2178 &  94.5 & 1.80 & 17.4 & 62.9 \\[3pt]
J124410$+$172104 & 0.5504 &  21.2 & 1.34 & 31.6 & 20.1 \\[3pt]
J130112$+$590206 & 0.1967 & 135.4 & 1.87 & 80.7 & 39.7 \\[3pt]
J131956$+$272808 & 0.6610 & 103.8 & 1.45 & 65.8 & 86.6 \\[3pt]
J132222$+$464546 & 0.2142 &  38.5 & 2.33 & 57.8 & 13.8 \\[3pt]
J134251$-$005345 & 0.2270 &  35.2 & 1.86 &  0.1 & 13.1 \\[3pt]
J135704$+$191907 & 0.4592 &  45.4 & 1.40 & 24.7 & 64.2 \\[3pt]
J155504$+$362847 & 0.1893 &  33.4 & 1.69 & 51.8 & 47.0 \\[3pt]
J213135$-$120704 & 0.4300 &  48.4 & 2.06 & 48.3 & 14.9 \\[3pt]
J225357$+$160853 & 0.3529 & 203.1 & 1.30 & 36.7 & 88.7 \\[-5pt]

\enddata
\end{deluxetable}   
%%%%%%%%%%%%%%%%%%%%%

We have modified the Kacprzak15 sample slightly in a few cases. The
sample we use is summarized in Table~\ref{tab:galprops}, and the
changes are as follows. The J121920 absorber--galaxy pair is excluded
here due to a highly uncertain azimuthal angle measurement. We
calculated galaxy colors for two additional galaxies using magnitudes
obtained from NED\footnote{\href{
    https://ned.ipac.caltech.edu/}{https://ned.ipac.caltech.edu/}}:
J100402 ($u-r=1.00$) and J111908 ($B-K=2.21$). Lastly, the colors
quoted by Kacprzak15 for the J045608 galaxies are transposed in their
Table 2; the values are published correctly in \citet{magiicat1}.

After followup observations with ESI/Keck spectra, we found that the
impact parameters of the two J1233$-$0031 galaxies ($z_{\rm gal} =
0.4174$ and $z_{\rm gal}=0.3185$) reported by \citet{werk12} appear to
be transposed. The $z_{\rm gal}=0.3185$ galaxy, which is included
here, is located at $D=85$~kpc rather than $D=31$~kpc. This error is
propagated in Kacprzak15, and the reported inclination and azimuthal
angle measurements should be $i = 38.7^{\circ}$ and
$\Phi=17.0^{\circ}$, respectively. This, however, does not
significantly change their results. We also later found that one of
the galaxies listed in the Kacprzak15 sample is located in a group
environment: Q0122$-$003, $z_{\rm gal}=0.3787$, with a neighboring
galaxy at $z_{\rm gal}=0.3792$. It is also listed in the {\magiicat}
sample as an isolated galaxy, though we do not have a HIRES or UVES
quasar spectrum for the associated absorber so it is not included in
the {\magiicat} IV or V analyses. For the work presented here, we do
not use this galaxy and have moved it to a group sample for later
analysis.

Each of the 29 galaxies in our sample was imaged with WFPC2, WFC3, or
ACS on the {\it Hubble Space Telescope (HST)} and their morphological
properties were modeled using GIM2D \citep{simard02}. Full details of
the galaxies and their modeling are described in Kacprzak15. We define
an inclination of $i=0^{\circ}$ as face-on and $i=90^{\circ}$ as
edge-on. An azimuthal angle of $\Phi=0^{\circ}$ indicates that the
background quasar sightline is aligned with the projected galaxy major
axis, and $\Phi=90^{\circ}$ indicates the sightline is aligned with
the projected galaxy minor axis.

Galaxy $u-r$ colors from Kacprzak15, plus an additional $u-r$ color
obtained from NED, were converted to $B-K$ colors following similar
methods described in \citet[][{\magiicat} I]{magiicat1} for direct
comparison to the {\MgII} sample studied in {\magiicat} IV and V,
which uses $B-K$ colors. Colors for each galaxy spectral energy
distribution (SED) were calculated and we obtained a linear
least-squares fit of $(B-K)=1.10(u-r)+0.113$ to the galaxy SED
colors. We then applied this relation to the $u-r$ colors to have a
uniform set of $B-K$ colors, and the new values are listed in
Table~\ref{tab:galprops}.

In order to examine the absorber kinematics for galaxies as a function
of different baryon cycle processes and star formation rates, we form
various subsamples by using the median galaxy property values of the
sample. For the orientation measurements, these values are $\langle i
\rangle = 51^{\circ}$ and $\langle \Phi \rangle = 45^{\circ}$. Using
the average inclination of galaxies in the universe, as was done in
{\magiicat} V, is not feasible here because the subsample sizes with
that cut would not be balanced. The uncertainties on the orientation
measurements are small enough such that only one galaxy could
potentially shift from being assigned to the minor axis subsample to
being assigned to the major axis subsample, and only two galaxies
could shift from face-on to edge-on.

The median galaxy color of the sample is $\langle B-K \rangle
=1.66$. This value is used to compare between galaxies that are more
likely to be star-forming or passive rather than with galaxy
morphological types or with the color bimodality of galaxies in the
universe. Comparing the absorption associated with blue galaxies to
that associated with red galaxies is important because previous work
has shown that the equivalent widths (which depend on the velocity
width and column density of the gas) of low-ionization {\MgII}
absorbers depend on some measure of the star formation rate
\citep[e.g.,][]{zibetti07, bordoloi11,
  rubin-winds14}. \citet{tumlinson11} found that {\OVI}, which is the
focus here, also depends on the SFR of the host galaxy, where
star-forming galaxies nearly always have detected absorption and
passive galaxies rarely have detected absorption. Since we do not
currently have star formation rates, we rely on galaxy color as a
proxy for comparison to the Tumlinson sample.

The cut used here, $\langle B-K \rangle = 1.66$, is roughly consistent
with the boundary between star-forming {\OVI} absorbing galaxies and
passive {\OVI} non-absorbing galaxies. Kacprzak15 show that the
boundary between mostly {\OVI} absorbing and mostly {\OVI}
non-absorbing galaxies is roughly at $B-K =1.6$ (same for
$u-r$). While Kacprzak15 do not have the star formation rates of the
galaxies, this cut is consistent with \citet{tumlinson11}.

The subsample cuts, median redshift, and number of galaxies in each
subsample are listed in Table~\ref{tab:v50}. We also list the
subsample cuts for the {\MgII} subsamples from {\magiicat} IV and V
for comparison.

%%%%%%%%%%%%%%%%%%%%%%%%%%%%%%%%%%%%%%%%%%%%%%%%%%%%%%%%%%%%%%%%%%%%%%%%%%%%%%%
% Figure 1
\begin{figure*}[ht]
\includegraphics[width=\linewidth]{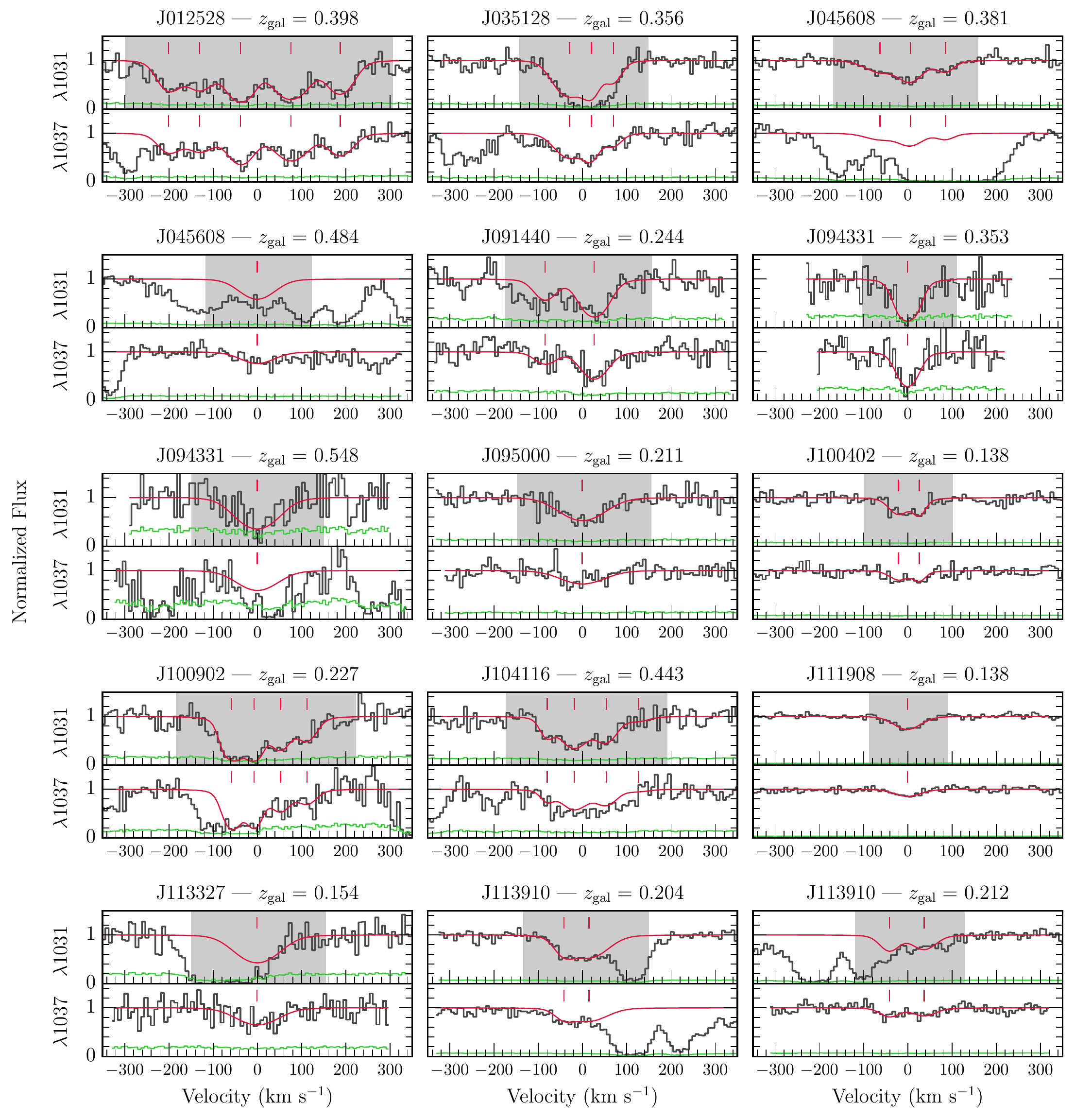}
\caption[]{{\OVIdblt} doublet absorption profiles and fits for each
  absorber--galaxy pair in the sample. In each panel pair, the
  $\lambda 1031$ line is plotted on top, and the $\lambda 1037$ line
  on bottom. The quasar spectrum is plotted as the black histogram,
  the uncertainty on the spectrum is the green line, and the fit to
  the data is plotted as the red line. Vertical red ticks at the top
  of each panel show the central velocity of each Voigt profile
  component fitted to the data. Gray shaded regions in the top panels
  indicate the velocity range of the absorbers, defined in
  Section~\ref{sec:spectra}. For the TPCF calculations, we use only
  those pixels located within these shaded regions. Velocity zero
  points are defined as the optical depth-weighted median of
  absorption.}
\end{figure*}

\addtocounter{figure}{-1}

\begin{figure*}[ht]
\includegraphics[width=\linewidth]{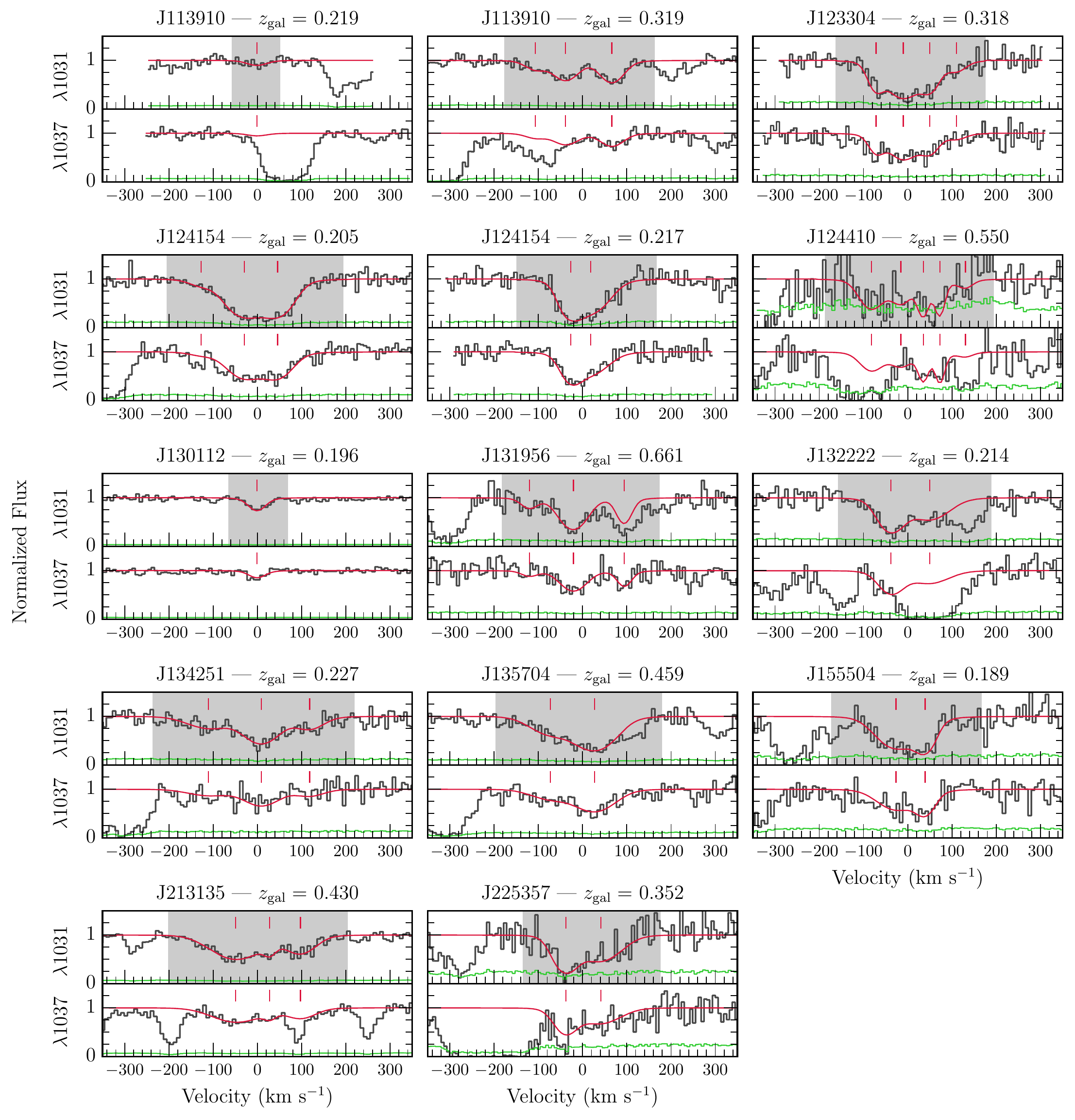}
\caption[]{(Continued)}
\label{fig:spec1}
\end{figure*}
%%%%%%%%%%%%%%%%%%%%%%%%%%%%%%%%%%%%%%%%%%%%%%%%%%%%%%%%%%%%%%%%%%%%%%%%%%%%%%%

There are no significant correlations between galaxy properties in the
sample. There are also no biases with azimuthal angle or inclination;
a one-dimensional Kolmogorov--Smirnov (KS) test reports that the
azimuthal angles and inclinations of the galaxies are consistent with
unbiased samples at the $0.6\sigma$ and $2.6\sigma$ levels,
respectively.

\subsection{Quasar Spectra}
\label{sec:spectra}

The galaxies described in the previous section are associated with
absorption in 23 quasars. Each quasar has a medium resolution
($R\sim20$,000, FWHM $\sim18${\kms}) spectrum from COS/{\it HST},
which covers the detected {\OVIdblt} doublet at the redshifts of the
targeted galaxies. Full details of the spectra, their reduction, and
the modeling of absorption are listed in Kacprzak15. Each {\OVIdblt}
doublet was Voigt profile (VP) fitted, simultaneously when possible to
account for blends, using
VPFIT.\footnote{\href{http://www.ast.cam.ac.uk/~rfc/vpfit.html}
  {http://www.ast.cam.ac.uk/$\sim$rfc/vpfit.html}} Velocity zero
points (i.e., $z_{\rm abs}$) were defined as the median velocity of
the optical depth distribution of absorption for the {\OVI}~$\lambda
1031$ line. The {\OVI} absorption doublets for the absorber--galaxy
pairs are plotted in Figure~\ref{fig:spec1}.

The velocity range of each absorber was determined by finding the
velocity or wavelength at which the VP model spectrum (rather than the
actual spectrum) decreases by 1\% from the continuum level. Using the
VP model to define the velocity ranges is necessary since several
{\OVI} absorbers are blended with other ions and the spectra are more
noisy than the HIRES or UVES spectra for {\MgII}. For our analysis, we
use only those pixels within these velocity bounds, which are plotted
as gray shaded regions in the $\lambda 1031$ panels of
Figure~\ref{fig:spec1}.

This method is simpler than the method used for {\MgII} because the
{\OVI} absorbers have a less complex absorption profile; the {\OVI}
absorbers generally consist of a single broad ``kinematic subsystem''
while {\MgII} may be composed of multiple kinematic subsystems
\citep[e.g.,][{\magiicat} IV, V]{cv01}. Only one {\OVI} absorber
(J121920) has two kinematic subsystems, but we exclude this absorber
from our analysis because the associated galaxy has a highly uncertain
azimuthal angle measurement. When we use this velocity range
determination method on the {\MgII} absorbers from {\magiicat} IV and
V, we find comparable, or more conservative ranges (i.e., velocities
closer to $v=0$~{\kms}) compared to our previous method in nearly all
cases. Thus, the different methods do not produce drastically
different results, and our simpler method may actually slightly
underestimate the velocity extents of the {\OVI} kinematics compared
to {\MgII}.

\subsection{The {\MgII} Comparison Sample}

We compare the {\OVI} absorber kinematics to the {\MgII} absorber
kinematics published in {\magiicat} IV and V. Here we briefly describe
the {\MgII} sample and refer the reader to the {\magiicat} series
papers for further details \citep[][]{magiicat2, magiicat1, magiicat5,
  magiicat4, magiicat3}.

We use a subset of 30 absorber--galaxy pairs with {\MgII} absorption
from the {\MgII} Absorber--Galaxy Catalog ({\magiicat}). All of these
galaxies have spectroscopic redshifts ($0.3 < z_{\rm gal} < 1.0$,
$\langle z_{\rm gal} \rangle = 0.656$), $B-K$ colors ($\langle B-K
\rangle = 1.4$), and {\MgII} absorption detected in high-resolution
background quasar spectra (HIRES/Keck or UVES/VLT) within a projected
distance of $D=200$~kpc ($\langle D \rangle = 40$~kpc). Additionally,
all galaxies have {\it HST} images with which the orientations have
been measured using GIM2D \citep{kcems11, kcn12}. Of the 30 {\MgII}
absorber--galaxy pairs, only seven have associated {\OVI} absorption
and overlap with the {\OVI} sample presented in
Section~\ref{sec:galprops}. Note that the absorber--galaxy pairs in
this {\magiicat} sample have higher redshifts, bluer colors, and
smaller impact parameters on average than our main {\OVI} sample.

Subsamples for the {\MgII} sample are determined by the median galaxy
color of $\langle B-K \rangle = 1.4$, as well as orientation
measurements of $i = 57^{\circ}$ (the mean inclination of galaxies in
the universe), and $\Phi = 45^{\circ}$. These values and the subsample
sizes are tabulated in Table~\ref{tab:v50} and are further described
in {\magiicat} IV and V.

The absorption kinematics for the {\MgII} sample have already been
fully analyzed in {\magiicat} IV and V. We present these kinematics
here for comparison, but do not present any new results.

%%%%%%%%%%%%%%%%%%%%%%%%%%%%%%%%%%%%%%%%%%%%%%%%%%%%%%%%%%%%%%%%%%%%%%%%%%%%%%%
% Figure 2
\begin{figure*}[ht]
\includegraphics[width=\linewidth]{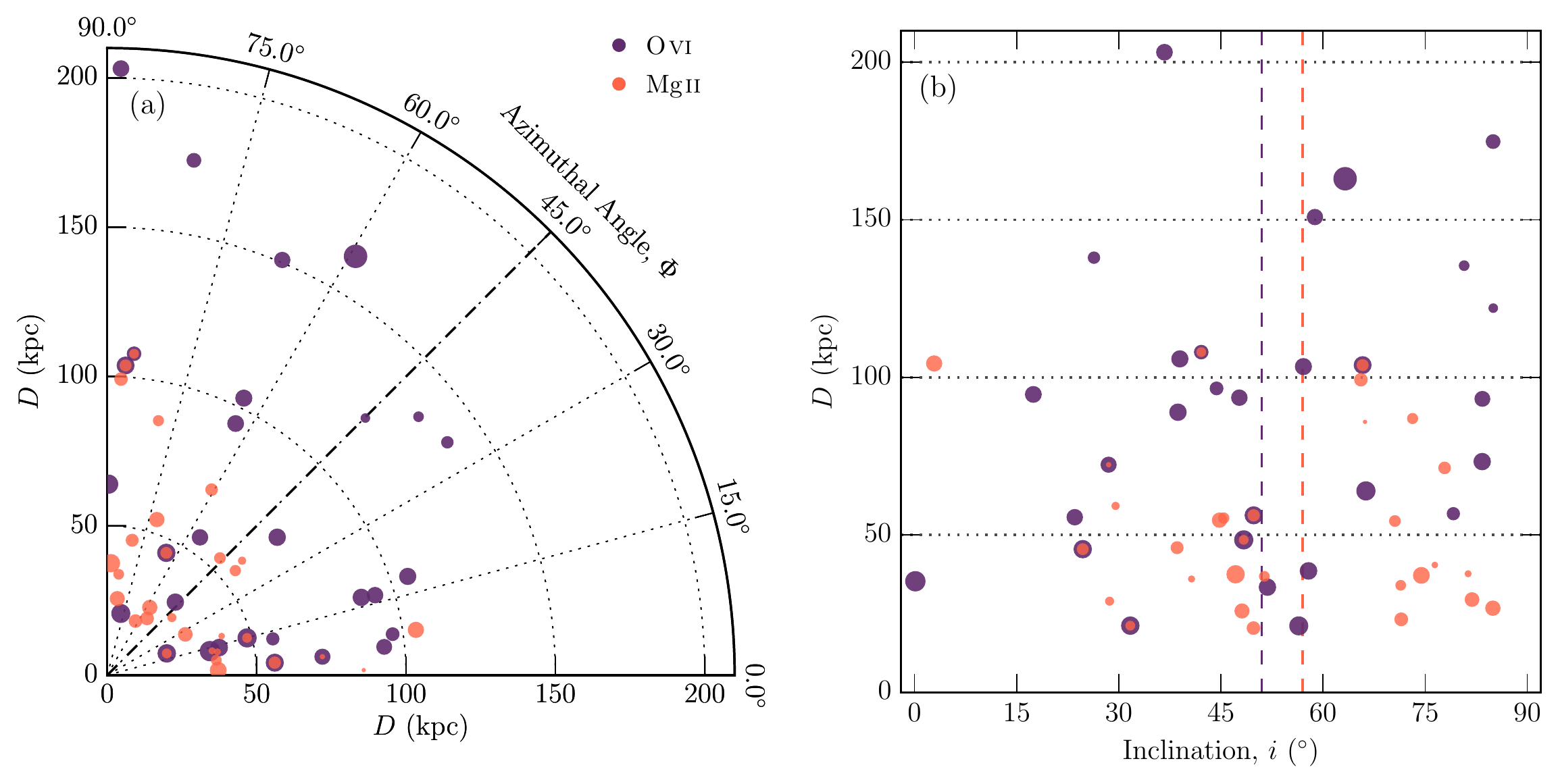}
\caption[]{Galaxy orientation measures as a function of impact
  parameter for both {\OVI} and {\MgII} samples. Point sizes represent
  the velocity spread of the absorbers, with larger points indicating
  larger velocity spreads. The seven overlapping {\OVI} and {\MgII}
  points indicate absorber--galaxy pairs for which we have detected
  both ions. (a) Azimuthal angle versus impact parameter for both the
  {\OVI} and {\MgII} samples. Points represent the location of the
  quasar sightline, where the foreground galaxy in each case is
  located at $D=0$~kpc and is aligned such that the major axis has
  $\Phi=0^{\circ}$ and the minor axis has $\Phi=90^{\circ}$. The
  dashed line at $\Phi=45^{\circ}$ indicates the value by which we
  slice the samples into major and minor axis subsamples. (b)
  Inclination versus impact parameter for both ion samples. Vertical
  dashed lines indicate the inclinations by which the samples are
  sliced, with $i=51^{\circ}$ for {\OVI} and $i=57^{\circ}$ for
  {\MgII}. {\OVI} absorbers tend to be located further from the
  galaxy, especially along the minor axis and for edge-on galaxies.}
\label{fig:sample}
\end{figure*}
%%%%%%%%%%%%%%%%%%%%%%%%%%%%%%%%%%%%%%%%%%%%%%%%%%%%%%%%%%%%%%%%%%%%%%%%%%%%%%%

%%%%%%%%%%%%%%%%%%%%%%%%%%%%%%%%%%%%%%%%%%%%%%%%%%%%%%%%%%%%%%%%%%%%%%%%%%%%%%%
% Figure 3
\begin{figure}[ht]
\includegraphics[width=\linewidth]{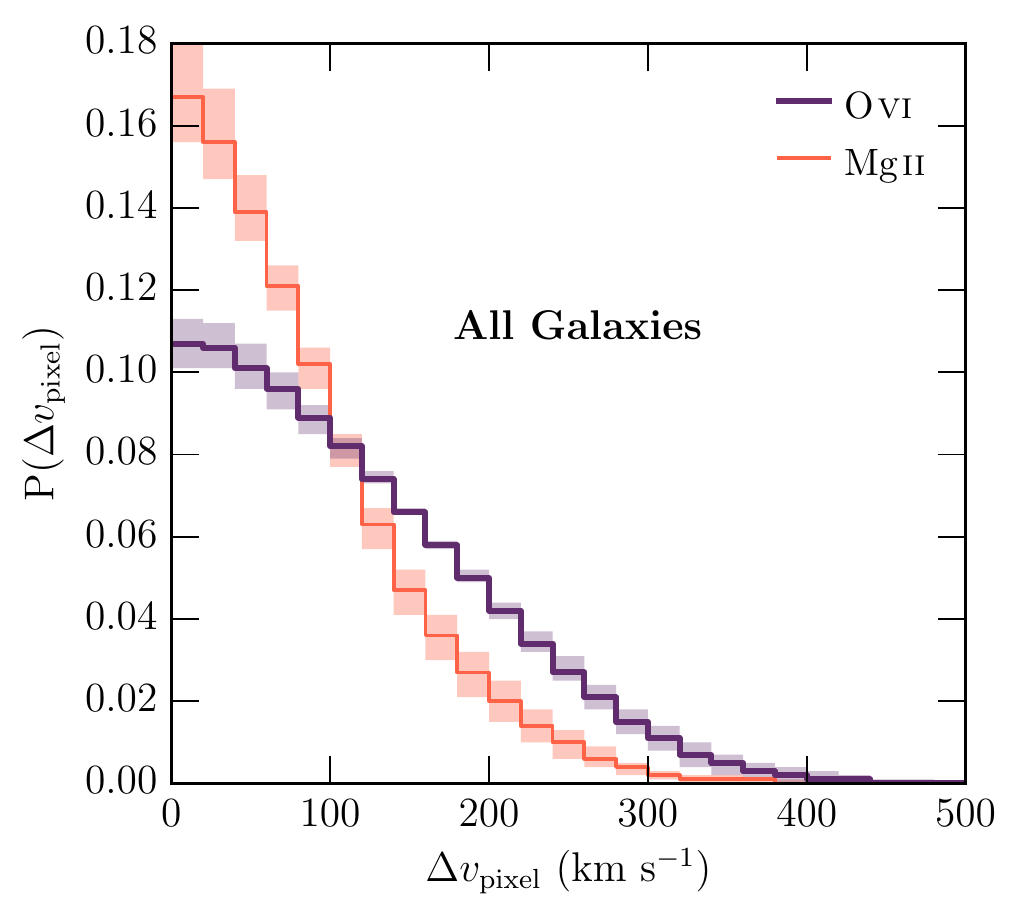}
\caption[]{Pixel-velocity TPCFs for the full samples of 29 {\OVI} and
  30 {\MgII} absorbers with the same binning for comparison between
  ions. The {\MgII} absorbers come from {\magiicat} IV and include
  only those absorber--galaxy pairs with galaxy colors and orientation
  measurements. The thick purple line and shading indicate {\OVI}
  TPCFs and uncertainties, respectively, while thin orange line and
  shading represent {\MgII}. {\OVI} absorbers have significantly
  larger velocity dispersions than {\MgII} absorbers.}
\label{fig:MgIIOVI}
\end{figure}
%%%%%%%%%%%%%%%%%%%%%%%%%%%%%%%%%%%%%%%%%%%%%%%%%%%%%%%%%%%%%%%%%%%%%%%%%%%%%%%

\subsection{Pixel-velocity TPCFs}
\label{sec:tpcfs}

To examine the absorber kinematics as a function of galaxy properties,
we use the pixel-velocity TPCF method described in detail in
{\magiicat} IV and V. A summary of the method follows.

The pixel-velocity TPCF is calculated by first obtaining the
velocities of all pixels within the velocity bounds of detected
absorption for a subsample. A velocity of $v=0$~{\kms} corresponds to
the optical depth-weighted median of absorption and defines the
absorption redshift, $z_{\rm abs}$. The pixel velocities for all
absorbers in a subsample are pooled together as if they came from a
single absorber hosted by a galaxy of a certain type, e.g., blue
galaxies probed along the projected minor axis. Velocity separations
between each pixel pair, without duplications, are then calculated for
this pool. The absolute value of these velocity separations are then
binned into 20~{\kms} wide bins, which is comparable to the FWHM of
COS/{\it HST}. The count in each bin is normalized by the total number
of pixel velocity pairs in the subsample to account for varying
subsamples sizes when comparing between subsamples. The pixel-velocity
TPCF is thus a probability distribution function and provides a
statistical view of the absorber velocity dispersion for a given
galaxy subsample.

Uncertainties in the TPCFs are calculated using a bootstrap analysis
with 100 realizations. The uncertainties reported are $1\sigma$
deviations from the mean of the bootstrap realizations, which allow
for asymmetrical uncertainties around the true TPCFs. 

To compare TPCFs between subsamples, we perform a chi-squared test,
taking into account the uncertainties in the TPCFs. We also report
{\vfifty} and {\vninety} values, i.e., the velocity separation within
which 50\% and 90\% of the area under the TPCF curve is contained, for
each TPCF to help describe where two TPCFs differ in more detail than
the chi-squared test provides. These values are tabulated in
Table~\ref{tab:v50}. The uncertainties on {\vfifty} and {\vninety} are
calculated from the bootstrap realizations, similar to the TPCF
uncertainties.

The bin sizes for the TPCFs presented here are twice as large as the
TPCFs presented with {\MgII} ({\magiicat} IV and V) due to a coarser
spectral resolution in the COS spectra compared to HIRES spectra. To
test the effect that changing the TPCF bin sizes had, we reran the
{\MgII} TPCFs from {\magiicat} IV and V with the 20~{\kms} bin
widths. With larger bin sizes, the general {\MgII} TPCF results
remained unchanged. These coarser {\MgII} TPCFs are presented with the
{\OVI} TPCFs for comparison between ions.

\section{Results}
\label{sec:results}

\subsection{Full Sample}

\subsubsection{Sample Distribution}

The distribution of both {\OVI} and {\MgII} absorbers as a function of
their orientation relative to the host galaxy is plotted in
Figure~\ref{fig:sample}. Panel (a) presents the azimuthal angle versus
impact parameter of each absorber--galaxy pair. The galaxy for each
pair is located at $D=0$~kpc, with the major axis aligned with
$\Phi=0^{\circ}$. Points represent the location of the background
quasar sightline. Point sizes indicate the velocity width of
absorption, i.e., the difference between the extremes of the gray
shaded regions in Figure~\ref{fig:spec1}. Point sizes can be compared
between ions as both are normalized to the maximum velocity spread of
the {\OVI} absorbers.

There is an overlap of seven absorber--galaxy pairs between the
{\MgII} and {\OVI} samples. These are presented as orange points on
top of purple. We note that although the overlapping points (the
differing ions) are plotted in the same locations, their $z_{\rm abs}$
values can differ by up to roughly $110$~\kms \citep[for the {\MgII}
  profiles, see][]{kcems11}. The smallest velocity separation between
the $z_{\rm abs}$ for the two ions is 27~{\kms}. Given that $z_{\rm
  abs}$ is the optical depth weighted median of absorption, this
indicates that the absorption is distributed differently along the
line of sight between the two ions.

The azimuthal angle behavior of the samples discussed in detail by
\citet{kcn12} ({\MgII}) and Kacprzak15 ({\OVI}) appears to be present
in this plot, though the trend is less obvious here. This is likely
because we only present a subset of the {\MgII} sample studied by
\citet{kcn12}. There is a population of absorbers for both ions within
$\Phi \sim 15^{\circ}$ of the major axis, a slight gap, and then
another, larger population at $\Phi>35^{\circ}$. Interestingly, the
group of three {\OVI} absorbers located at $D\sim 140$~kpc and
$30^{\circ} < \Phi < 45^{\circ}$ appear to have smaller point sizes
than the rest of the {\OVI} absorbers. As shown in Kacprzak15, the
frequency of ``non-detections'' in {\OVI} ($W_r(1031)<0.1$~{\AA}) is
largest between $30^{\circ} < \Phi < 60^{\circ}$. In fact, these three
points have $W_r(1031)<0.1$~{\AA}, the only absorbers with equivalent
widths this low in the sample presented here. We refrain from
investigating the azimuthal angle distribution preferences of the
absorbers further as these were examined previously.

Figure~\ref{fig:sample}(b) presents the impact parameter, $D$, as a
function of inclination, $i$, for the {\MgII} and {\OVI}
samples. Point sizes again represent the velocity width of
absorption. The vertical dashed lines represent the inclinations by
which we slice the sample into ``face-on'' ($i\sim 0^{\circ}$) and
``edge-on'' ($i\sim 90^{\circ}$) subsamples. For the {\MgII} sample,
we used the mean inclination of galaxies in the universe, whereas here
we use the median inclination of the {\OVI} absorbers to even out
subsample sizes. The {\MgII} absorbers appear to have a larger
variation in their point sizes (velocity spreads) than {\OVI}, though
this is subtle. Also difficult to discern is the dependence of the
point sizes on $\Phi$, $i$, and $D$. We use the pixel-velocity TPCFs
to examine these differences in more detail below.

\subsubsection{Pixel-velocity TPCFs}

As a first comparison of the kinematics between the different ions,
the TPCF for the full sample of {\OVI} absorbers is plotted as a thick
purple line with shading representing errors in
Figure~\ref{fig:MgIIOVI}. The full sample of {\MgII} absorbers with
the same binning is plotted as the thin orange line with shading
representing errors. The velocity dispersion of {\OVI} tends to be
large, with pixel velocity separations up to $\sim 400$~{\kms}. This
is in contrast to the more narrow {\MgII} TPCF. Compared to {\MgII},
{\vfifty} and {\vninety} for {\OVI} are roughly 50\% and 40\% larger,
respectively.

It is important to keep in mind that our {\OVI} sample of galaxies is
located at lower redshifts ($\langle z_{\rm gal} \rangle = 0.244$), is
probed at greater distances on average ($\langle D \rangle =
93.2$~kpc, as shown in Figure~\ref{fig:sample}), and has redder colors
($\langle B-K \rangle = 1.66$) than the {\MgII} galaxies. We discuss
these differences in Section~\ref{sec:discussion}. For easy comparison
between the two ions, the subsample cuts and sizes for each ion are
listed in Table~\ref{tab:v50} for the rest of the presented results.

%%%%%%%%%%%%%%%%%%%%%
\begin{deluxetable*}{lccccllcccc}
\tablecolumns{10} 
\tablewidth{0pt} 
\tablecaption{TPCF {\vfifty} and {\vninety}
  Measurements \label{tab:v50}}
\tablehead{
  \colhead{} &
  \multicolumn{6}{c}{{\OVI}} &
  \colhead{} &
  \multicolumn{3}{c}{{\MgII}\tablenotemark{a}} \\
  \cline{2-7} \cline{9-11} \\[-5pt]
  \colhead{Sample} &
  \colhead{Cut} &
  \colhead{Cut} &
  \colhead{$\langle z_{\rm gal} \rangle$} &
  \colhead{\# Gals} &
  \colhead{{\vfifty}\tablenotemark{b}}  &
  \colhead{{\vninety}\tablenotemark{b}} &
  \colhead{} &
  \colhead{Cut} &
  \colhead{Cut} &
  \colhead{\# Gals} 
}
\startdata

\cutinhead{Figure~\ref{fig:MgIIOVI}}\\[-3pt]
All {\MgII} Absorbers\tablenotemark{a} & ${\cdots}$ & ${\cdots}$ & 0.656~\tablenotemark{c}  & 30~\tablenotemark{c}  & \phn$ 66_{- 6}^{+ 5}$~\tablenotemark{c}  & $172_{-17}^{+13}$~\tablenotemark{c} && ${\cdots}$ & ${\cdots}$ & ${\cdots}$ \\[3pt]
All {\OVI} Absorbers  & ${\cdots}$ & ${\cdots}$ & 0.244\phn & 29\phn & $100_{- 6}^{+ 5}$ & $235_{-16}^{+14}$ && ${\cdots}$ & ${\cdots}$ & ${\cdots}$ \\[3pt]

\cutinhead{Figure~\ref{fig:BKPA}}\\[-3pt]
Blue--Major Axis & {\blue} & {\major} & 0.319 &  7 & \phn$ 90_{- 8}^{+ 5}$ & $210_{-20}^{+11}$ && $B-K<1.4$    & $\Phi<45^{\circ}$      &  5 \\[3pt]
Blue--Minor Axis & {\blue} & {\minor} & 0.459 &  7 & \phn$ 96_{- 6}^{+ 5}$ & $225_{-14}^{+10}$ && $B-K<1.4$    & $\Phi \geq 45^{\circ}$ & 10 \\[3pt]
Red--Major Axis  & {\red}  & {\major} & 0.214 &  9 & \phn$ 99_{- 8}^{+ 9}$ & $231_{-18}^{+21}$ && $B-K\geq1.4$ & $\Phi<45^{\circ}$      & 10 \\[3pt]
Red--Minor Axis  & {\red}  & {\minor} & 0.215 &  6 &     $119_{-21}^{+17}$ & $280_{-51}^{+37}$ && $B-K\geq1.4$ & $\Phi \geq 45^{\circ}$ &  5 \\[3pt]

\cutinhead{Figure~\ref{fig:BKincl}}\\[-3pt]
Blue--Face-on & {\blue} & {\face}  & 0.353 &  9 & \phn$ 91_{- 5}^{+ 4}$ & $212_{-13}^{+ 9}$ && $B-K<1.4$    & $i<57^{\circ}$      &  8 \\[3pt]
Blue--Edge-on & {\blue} & {\edge}  & 0.319 &  5 & \phn$ 98_{- 8}^{+ 7}$ & $227_{-17}^{+16}$ && $B-K<1.4$    & $i \geq 57^{\circ}$ &  7 \\[3pt]
Red--Face-on  & {\red}  & {\face}  & 0.222 &  6 &     $106_{-11}^{+ 7}$ & $246_{-24}^{+16}$ && $B-K\geq1.4$ & $i<57^{\circ}$      &  9 \\[3pt]
Red--Edge-on  & {\red}  & {\edge}  & 0.212 &  9 &     $109_{-16}^{+17}$ & $262_{-40}^{+38}$ && $B-K\geq1.4$ & $i \geq 57^{\circ}$ &  6 \\[3pt]

\cutinhead{Figure~\ref{fig:PAincl}}\\[-3pt]
Face-on--Major Axis & {\face} & {\major} & 0.336 & 10 &     $100_{- 9}^{+ 6}$ & $234_{-20}^{+14}$ && $i<57^{\circ}$      & $\Phi<45^{\circ}$      & 10 \\[3pt]
Face-on--Minor Axis & {\face} & {\minor} & 0.353 &  5 & \phn$ 90_{- 4}^{+ 5}$ & $210_{-10}^{+11}$ && $i<57^{\circ}$      & $\Phi \geq 45^{\circ}$ &  7 \\[3pt]
Edge-on--Major Axis & {\edge} & {\major} & 0.209 &  6 & \phn$ 80_{-15}^{+10}$ & $187_{-32}^{+20}$ && $i \geq 57^{\circ}$ & $\Phi<45^{\circ}$      &  5 \\[3pt]
Edge-on--Minor Axis & {\edge} & {\minor} & 0.305 &  8 &     $116_{-15}^{+12}$ & $272_{-37}^{+28}$ && $i \geq 57^{\circ}$ & $\Phi \geq 45^{\circ}$ &  8 \\[-5pt]

\enddata
\tablenotetext{a}{The {\MgII} data, analysis, results, and conclusions
  are published in \citet{magiicat5} and \citet{magiicat4}, and
  references therein}
\tablenotetext{b}{{\kms}}
\tablenotetext{c}{Values listed are for {\MgII} absorbers}
\end{deluxetable*}
%%%%%%%%%%%%%%%%%%%%%

%%%%%%%%%%%%%%%%%%%%%%%%%%%%%%%%%%%%%%%%%%%%%%%%%%%%%%%%%%%%%%%%%%%%%%%%%%%%%%%
% Figure 4
\begin{figure}[ht]
\includegraphics[width=\linewidth]{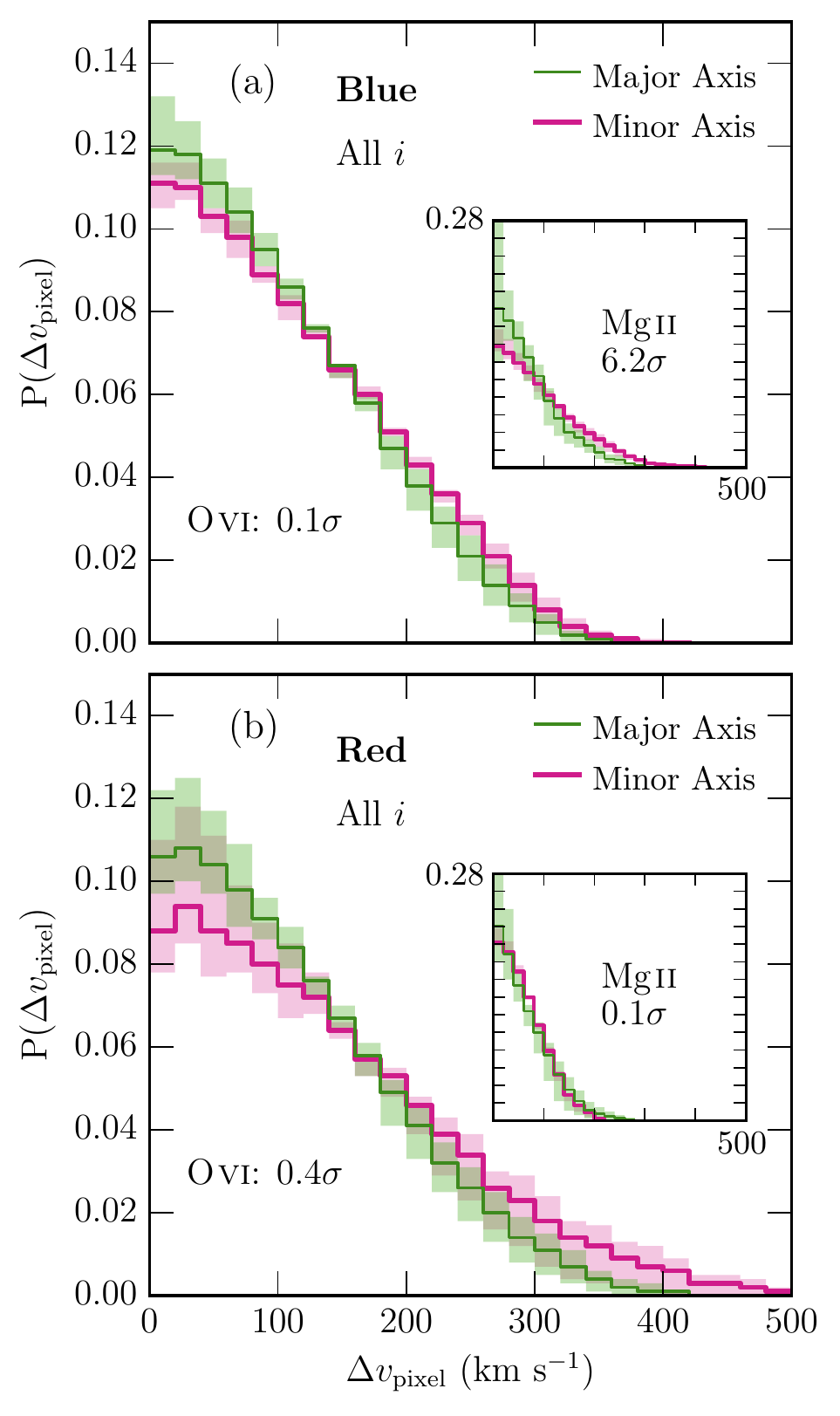}
\caption[]{Pixel-velocity TPCFs for (a) blue galaxies and (b) red
  galaxies probed along the projected major and minor axis. The TPCFs
  in the main panels represent {\OVI}, while those in the insets are
  for {\MgII}. The {\MgII} TPCFs are published in {\magiicat} V, and
  are rebinned here for comparison. The inset axes range in velocity
  separation from 0~{\kms} to 500~{\kms} (the same as the main panel),
  while the probabilities range from 0 to 0.28. The {\OVI} TPCFs show
  no dependence on $B-K$ nor $\Phi$, with the highest chi-squared
  significance result of $0.9\sigma$ for every combination of
  TPCFs. The {\MgII} TPCFs have smaller velocity dispersions for every
  subsample except for the blue, minor axis subsample which has
  velocity dispersions more similar to the higher ionization gas.}
\label{fig:BKPA}
\end{figure}
%%%%%%%%%%%%%%%%%%%%%%%%%%%%%%%%%%%%%%%%%%%%%%%%%%%%%%%%%%%%%%%%%%%%%%%%%%%%%%%

%%%%%%%%%%%%%%%%%%%%%%%%%%%%%%%%%%%%%%%%%%%%%%%%%%%%%%%%%%%%%%%%%%%%%%%%%%%%%%%
% Figure 5
\begin{figure}[ht]
\includegraphics[width=\linewidth]{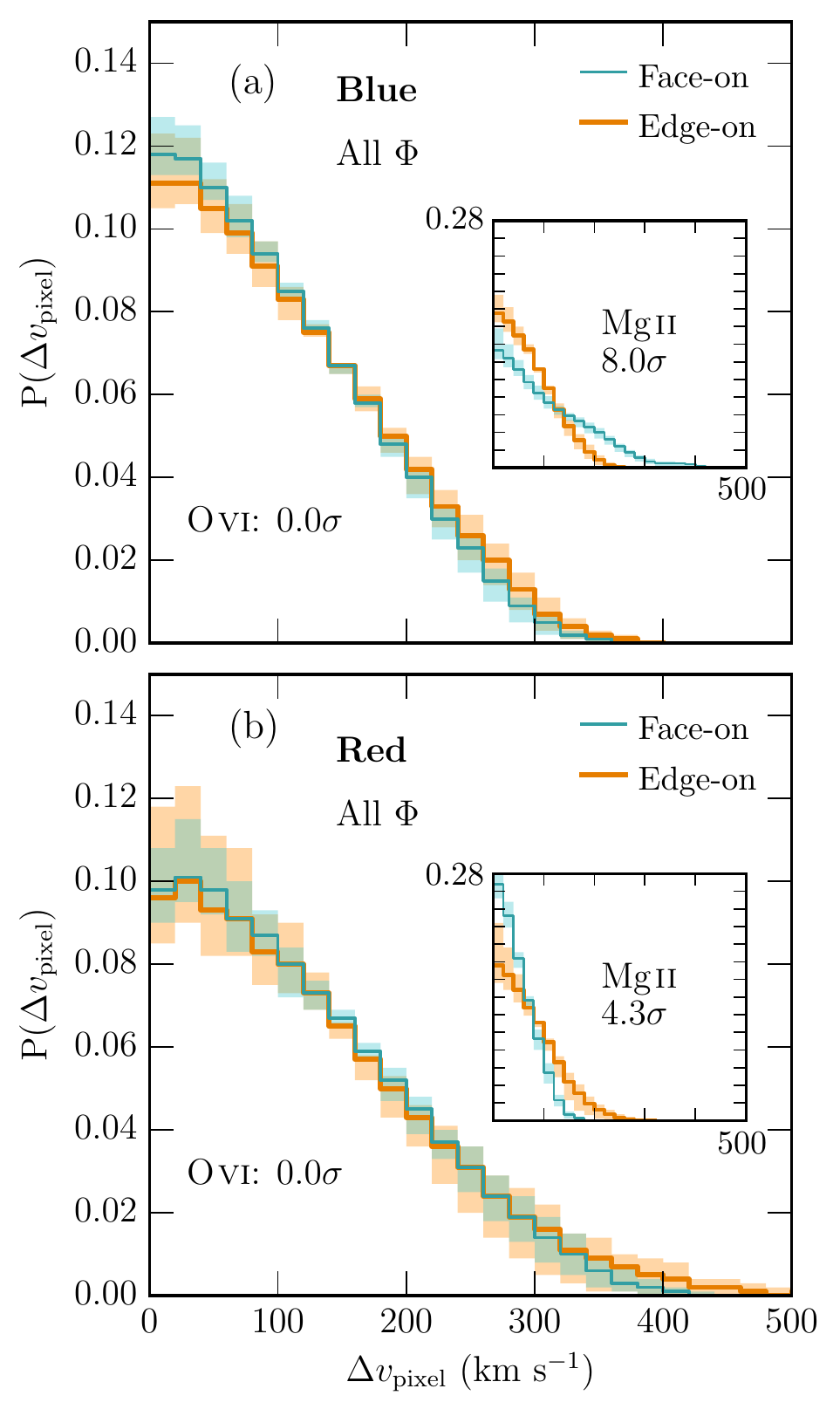}
\caption[]{Pixel-velocity TPCFs comparing face-on and edge-on
  orientations for (a) blue galaxies and (b) red galaxies. The {\OVI}
  and {\MgII} TPCFs are plotted as shown in Figure~\ref{fig:BKPA}. The
  {\MgII} TPCFs are published in {\magiicat} V, and are rebinned here
  for comparison.  There is no difference in the {\OVI} TPCFs for
  every subsample comparison, where the largest significance from a
  chi-squared test is $1.6\sigma$ for edge-on galaxies (not
  plotted). Only the blue, face-on subsample for the {\MgII} TPCFs has
  a comparable velocity dispersion to {\OVI}, the rest have smaller
  dispersions.}
\label{fig:BKincl}
\end{figure}
%%%%%%%%%%%%%%%%%%%%%%%%%%%%%%%%%%%%%%%%%%%%%%%%%%%%%%%%%%%%%%%%%%%%%%%%%%%%%%%

\subsection{Color and Orientation}

The TPCFs of blue galaxies (panel (a)) and red galaxies (panel (b))
probed along the projected major and minor axes are plotted in
Figure~\ref{fig:BKPA}. There are no differences in the {\OVI} TPCFs
for either panel ($0.1\sigma$ and $0.4\sigma$) and {\vfifty} and
{\vninety} for each subsample pair are all consistent within
uncertainties. For the subsample pairs not shown, the chi-squared
results are $0.1\sigma$ (major axis, blue vs. red galaxies) and
$0.4\sigma$ (minor axis, blue vs. red galaxies). The slightly (but not
significantly) larger velocity separation tail for red galaxies in
Figure~\ref{fig:BKPA}(b) is due to a single absorber with
$W_r(1031)=0.817$~\AA, which is an outlier in equivalent width for the
sample. This unusually strong {\OVI} equivalent width absorber was
studied in detail by \citet{muzahid15}, who associated the absorption
with a large-scale outflow. Removing this absorber does not change the
conclusions drawn from the TPCFs. The TPCFs of the {\MgII} absorbers
for the same subsamples ({\magiicat} V, inset panels) are
comparatively more narrow, with the exception of the blue, minor axis
subsample in panel (a), which has a velocity dispersion similar to the
{\OVI}.

For the TPCFs in Figure~\ref{fig:BKincl}, which compare face-on and
edge-on inclinations for blue galaxies (panel (a)) and red galaxies
(panel (b)), there are again no differences in the velocity
dispersions for each subsample pair ($\sim 0\sigma$ for all
pairs). The {\vfifty} and {\vninety} are also all consistent within
uncertainties, and the conclusions do not change when the large
equivalent width absorber is removed from the sample. In comparison,
while the {\MgII} TPCFs (published in {\magiicat} V, inset panels) for
edge-on subsamples are consistent within uncertainties, similar to the
behavior of {\OVI}, they have smaller velocity dispersions than
{\OVI}. While the {\MgII} TPCF for the red, face-on galaxy subsample
is also much more narrow than {\OVI}, the {\MgII} TPCF for the blue,
face-on galaxy subsample is comparable to {\OVI}.

These results indicate that the kinematics of {\OVI} absorbers do not
depend strongly on galaxy color or star formation activity for various
inclinations and azimuthal angles, in contrast to {\MgII} absorbers as
we found in {\magiicat} V.

%%%%%%%%%%%%%%%%%%%%%%%%%%%%%%%%%%%%%%%%%%%%%%%%%%%%%%%%%%%%%%%%%%%%%%%%%%%%%%%
% Figure 6
\begin{figure}[ht]
\includegraphics[width=\linewidth]{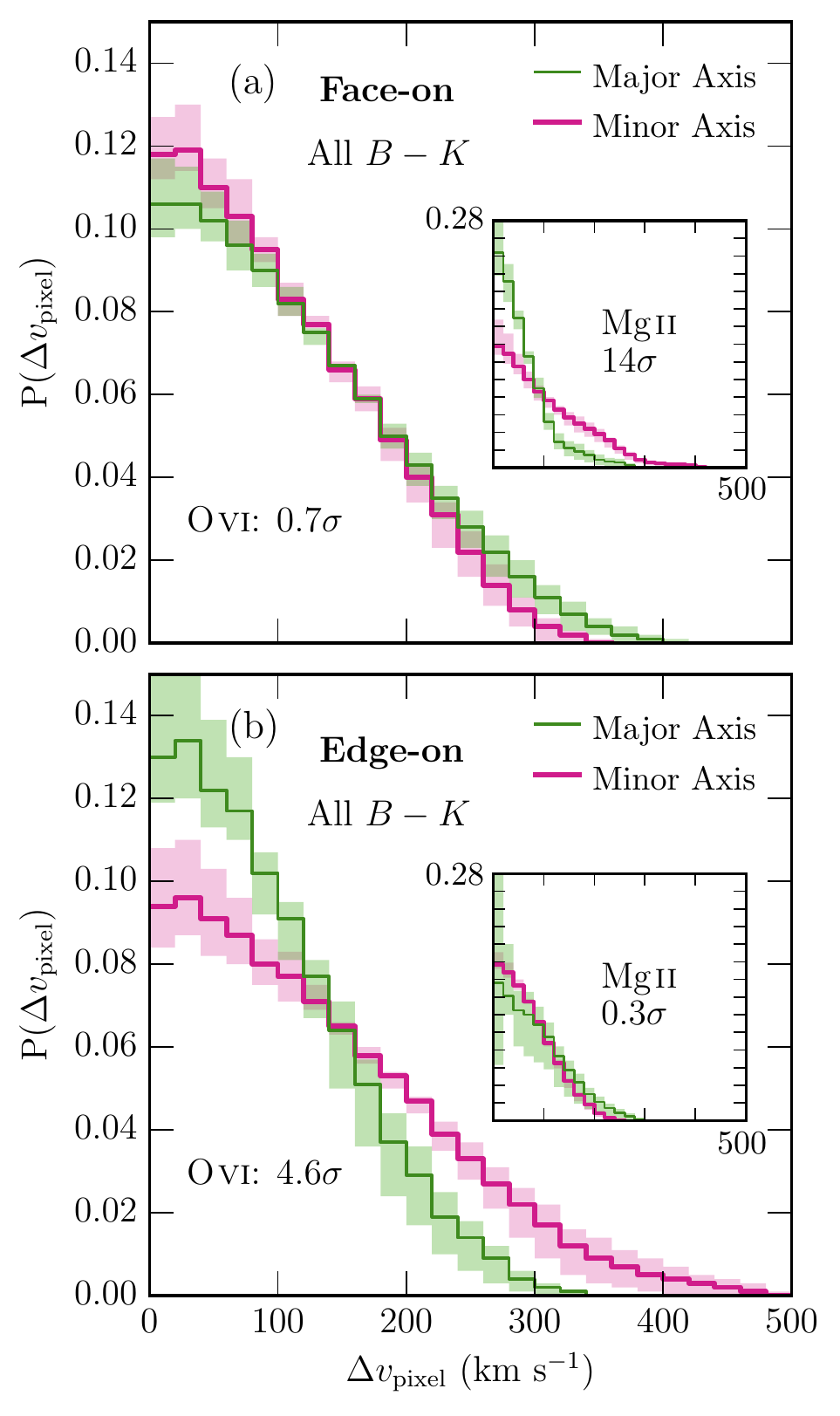}
\caption[]{Pixel-velocity TPCFs for (a) face-on galaxies and (b)
  edge-on galaxies probed along the projected major and minor
  axes. The TPCFs for the {\OVI} and {\MgII} subsamples are plotted
  similarly to those in Figure~\ref{fig:BKPA}. The {\MgII} TPCFs are
  published in {\magiicat} V, and are rebinned here for
  comparison. The {\OVI} TPCFs for face-on galaxies in panel (a) show
  no difference when the galaxy is probed along the major or minor
  axis ($0.7\sigma$). In panel (b) the {\OVI} TPCF for edge-on
  galaxies probed along the minor axis have larger velocity
  dispersions than those along the major axis, corresponding to a
  $4.6\sigma$ significance. This significance drops to $3.4\sigma$
  when the outlier in {\OVI} equivalent width is removed from the
  sample, and this remaining significant difference is due to a
  combination of small uncertainties on the minor axis subsample and
  the major axis being slightly less extended than the rest of the
  subsamples.}
\label{fig:PAincl}
\end{figure}
%%%%%%%%%%%%%%%%%%%%%%%%%%%%%%%%%%%%%%%%%%%%%%%%%%%%%%%%%%%%%%%%%%%%%%%%%%%%%%%

\subsection{Inclination and Azimuthal Angle}

For face-on galaxies in Figure~\ref{fig:PAincl}(a), we find no
differences ($0.7\sigma$) in the {\OVI} TPCFs for galaxies probed
along the projected major or minor axes. The {\vfifty} and {\vninety}
are consistent within uncertainties. This is in contrast to the
{\MgII} results from {\magiicat} V (inset panels), where face-on,
major axis galaxies host absorbers with much smaller velocity
dispersions. The face-on, minor axis {\MgII} TPCF is comparable to the
{\OVI} TPCF.

Conversely, we find significant differences in the {\OVI} TPCFs for
edge-on galaxies probed along the major and minor axes in
Figure~\ref{fig:PAincl}(b). Galaxies probed along the minor axis have
larger velocity dispersions than those probed along the major axis,
and this result is significant at the $4.6\sigma$ level. Both
{\vfifty} and {\vninety} for these samples are inconsistent within
uncertainties. A moderate portion of this difference is caused by the
$W_r(1031)=0.817$~\AA~absorber \citep[an edge-on, minor axis
  galaxy;][]{muzahid15}; however, removing the absorber from the
sample still results in a significance of $3.4\sigma$, though the
scatter on the edge-on, minor axis subsample is significantly
reduced. This result is in contrast to our previously published
results with {\MgII} ({\magiicat} V, inset panels), which have TPCFs
consistent within uncertainties. The edge-on, major axis subsample for
{\OVI} is comparable to the {\MgII} subsamples.

For face-on and edge-on galaxies probed along the major axis (not
plotted), we find an insignificant difference of $2.5\sigma$ where
face-on galaxies tend to have a slightly larger velocity spread than
edge-on galaxies. This behavior is opposite what we found in {\MgII}
({\magiicat} V), though the edge-on {\MgII} subsample has a velocity
dispersion comparable to the same {\OVI} subsample.

Finally, for galaxies probed along the minor axis (not plotted), we
find a significant difference ($4.0\sigma$) between the TPCFs of
face-on and edge-on galaxies, where absorbers hosted by edge-on
galaxies tend to have larger velocity spreads. However, the
significance decreases to $0.9\sigma$ when we remove the outlying
$W_r(1031)=0.8$~\AA~absorber. This behavior is also opposite what we
found with {\MgII}. As stated above, the {\MgII} face-on, minor axis
subsample is comparable to {\OVI}.

As we showed in the previous paragraphs, removing the outlying large
equivalent width absorber does not change the results in one
comparison (Figure~\ref{fig:PAincl}(b)), but removes the significant
difference in another (minor axis, face-on vs edge-on
subsamples). This results from the combination that the edge-on, major
axis subsample TPCF is slightly more narrow than the rest of the
subsamples, but the uncertainties are large enough to overlap with the
face-on subsamples, and the uncertainties on the edge-on, minor axis
subsample are reduced when the outlier is removed. Given this, we make
the assumption that there are no kinematic differences between these
subsamples when the outlying large equivalent width absorber is not
included. We also note that a larger sample size would be beneficial
in either reducing the uncertainties on the kinematics for the edge-on
subsamples along the major and minor axes, or showing more strongly
that there is a large variation in the velocity dispersions for these
subsamples.

\section{Discussion}
\label{sec:discussion}

As we have shown in the previous sections, the kinematics of {\OVI}
absorbers are similar regardless of galaxy color, azimuthal angle, and
inclination. Compared to our previously published {\MgII} TPCFs
({\magiicat} IV and V), the {\OVI} TPCFs are more extended and are
much less sensitive to the galaxy properties. In subsamples where we
expect outflows to dominate the absorption, {\MgII} and {\OVI} TPCFs
are comparable.

The larger velocity spreads for {\OVI} than {\MgII} for nearly all
subsamples may be explained if {\OVI} is collisionally ionized. In
this case, {\OVI} absorption profiles would be broader than
{\MgII}. However, the contrasts between the two ion samples still seem
a bit puzzling. If the two ions trace the same gas, then the relative
behavior of the TPCFs between the ions should be similar as they are
the result of the same baryon cycle processes, i.e., in
Figure~\ref{fig:BKincl}(a), the face-on subsample should be more
extended than the edge-on subsample for both {\OVI} and {\MgII}, but
this is not the case. If we assume that this scenario is true (we show
later that it likely is not), we can examine the several differences
between the two samples that may be contributing to our TPCF results.

The {\OVI} sample probes the CGM at larger impact parameters on
average ($\langle D \rangle = 86$~kpc) than the {\MgII} sample
($\langle D \rangle = 40$~kpc; see Figure~\ref{fig:sample};
{\magiicat} IV and V). Simulations by \citet{oppenheimer16} show an
age--radius anti-correlation with {\OVI} absorption, where the time
since the gas was ejected from the galaxy via outflows increases with
increasing radii. The lack of a kinematic dependence on orientation in
the {\OVI} sample TPCFs may then be a result of the absorbers being
located further away from the galaxy, and thus less dependent on the
current star formation activity than for the {\MgII} sample. This is
strengthened by the finding that the {\OVI} TPCFs show similar
behaviors for both blue and red galaxies.

The galaxies hosting {\OVI} absorbers tend to be redder than {\MgII}
host galaxies, with $\langle B-K \rangle =1.66$ and $\langle B-K
\rangle =1.4$, respectively. For {\MgII}, the kinematics of absorbers
around redder galaxies are less sensitive to the orientation at which
they are located than they are for blue galaxies (for example, see the
inset TPCFs in Figures~\ref{fig:BKPA} and \ref{fig:BKincl}; also see
{\magiicat} V). Additionally, redder galaxies tend to have lower star
formation rates and, consequently, are not expected to have active
outflows or accretion. Given this, we would expect less of an
orientation dependence for the {\OVI} absorbers because they are
redder galaxies on average than the {\MgII} sample, which does show an
orientation dependence.

The redshift distributions of the two samples are also
different. While the galaxies for the {\MgII} sample were located at a
median redshift of $\langle z_{\rm gal} \rangle = 0.656$ ($0.3<z_{\rm
  gal}<1.0$), the {\OVI} sample has a median of $\langle z_{\rm gal}
\rangle = 0.244$ ($0.1<z_{\rm gal}<0.66$). In this case, the lower
redshift galaxies have lower star formation rates (they tend to be
redder), and thus have a less active baryon cycle than at higher
redshifts. This would result in a weaker dependence of the absorber
kinematics on galaxy orientation for the lower redshift {\OVI}
sample. \citet{kacprzak11kin} reported the fraction of systems where
{\MgII} absorption velocities were in alignment with the galaxy
rotation direction. They found that this fraction decreases by a
factor of two from $z\sim 0.5$ to $z\sim0.1$. Furthermore, they report
a factor of three increase at $z\sim 0.1$ compared to $z\sim 0.5$ for
{\MgII} absorption spanning both sides of the galaxy systemic
velocity. It is possible that CGM kinematics may evolve with redshift;
however, we have yet to explore the {\MgII} TPCFs in the lower
redshift range probed by the {\OVI} sample because of the lack of
high-resolution spectra.

While the differences between the two samples listed above probably do
have some effect on our results, we do not think they dominate. The
discussion above mostly assumes that {\OVI} and {\MgII} trace the same
components of the CGM. However, it is more likely that, with their
differing ionization states and kinematics, the two ions trace
different components of the CGM. This is supported by both
observations \citep[e.g.,][]{werk13, muzahid15} and simulations
\citep[e.g.,][]{ford14, churchill15}. Even for the overlapping seven
galaxies in the {\MgII} and {\OVI} samples with absorption in both
ions, the two ions are offset in $z_{\rm abs}$ (defined as the optical
depth-weighted median of the absorption) by as little as
$\sim10$~{\kms} or as great as $\sim100$~{\kms}. These seven absorbers
show that the distribution of the gas is different in velocity space
between ions.

Using roughly the same sample we use here, Kacprzak15 showed that
{\OVI} absorption is preferentially distributed along the major and
minor axes of the host galaxies, with non-detected {\OVI} sightlines
primarily located between the major and minor axes. The authors
suggested that {\OVI} is not mixed throughout the CGM and is confined
into outflows along the minor axis and inflows or recycled gas along
the major axis. They also showed that the {\OVI} equivalent widths
were stronger along the projected minor axis than along the major
axis, suggesting that either the column densities, the velocity
spreads, or both are larger for absorbers located along the minor
axis. These results seem contrary to what we find here with the
kinematics.

Since the kinematics of the {\OVI} absorbers are roughly consistent
for all galaxy color and orientation combinations, this suggests that
{\OVI} may not trace different baryon cycle processes. In simulations,
\citet{ford14} found that {\OVI} traces gas that was ejected from the
galaxy by ancient outflows many Gyr prior to the current epoch, some
of which is likely reaccreting onto the galaxy at the time of the mock
observations. If this is the case, then the absorbing gas may have
plenty of time to mix and form a roughly kinematically uniform {\OVI}
halo at all locations about the galaxy. The sizes of {\OVI} absorbers
are also predicted to be large, on the order of tens to hundreds of
kiloparsecs from photoionization modeling \citep{lopez07, muzahid14,
  hussain15}, so any kinematic differences with galaxy orientation may
be wiped out by the large cloud sizes. This is in contrast to the
small, $\sim10$~pc cloud sizes expected for {\MgII}
\citep[e.g.,][]{weakII, crighton15}. A kinematically uniform {\OVI}
halo is therefore reasonable.

The combination of consistent absorber kinematics with orientation and
color, the low fraction of absorbing gas in intermediate azimuthal
angles from Kacprzak15, and absorption concentrated near the major and
minor axes may be the result of differing ionization conditions
throughout the CGM. {\OVI} absorbing gas that is located in the
intermediate azimuthal angles may be more susceptible to being ionized
out of the {\OVI} phase due to lower densities, resulting in lower
equivalent widths (or upper limits on the equivalent width) and lower
covering fractions. Conversely, higher gas densities due to outflows
and inflows along the minor and major axes, respectively, may shield
the oxygen from being ionized out of the {\OVI} ionization state and
provide more suitable conditions for {\OVI}. If so, higher ionization
phases of oxygen may show an orientation dependence in covering
fraction and equivalent width, with possibly a higher incidence of gas
in the intermediate azimuthal angles when compared to {\OVI}.

\section{Summary and Conclusions}
\label{sec:conclusions}

Using an absorption-selected sample of 29 galaxies ($0.13 < z_{\rm
  gal} < 0.66$) from Kacprzak15, we examined the velocity dispersion
of {\OVI} absorption as a function of galaxy color, inclination, and
azimuthal angle. Each absorber--galaxy pair was identified as part of
the ``Multiphase Galaxy Halos'' survey \citep[e.g.,
  Kacprzak15,][]{muzahid15, muzahid16} or obtained from the
literature. The galaxies were found within $D\sim 200$~kpc of a
background quasar sightline and have redshifts consistent with
detected {\OVI} absorption in {\it HST}/COS quasar spectra. Each
galaxy is isolated, with no nearby neighbors within 100~kpc and a
line-of-sight velocity separation of 500~{\kms}. Galaxies were modeled
using GIM2D to obtain their inclinations, morphologies, and the
azimuthal angle at which the background quasar probes the CGM relative
to the projected galaxy major axis. We use the pixel-velocity TPCF
method described in {\magiicat} IV and V for {\MgII} on our {\OVI}
absorbers, and compare the results between both {\MgII} from our
previous work and the {\OVI} here. Our findings include the following:

\begin{enumerate}[nolistsep]

  \item In general, the {\OVI} TPCFs are more extended than for
    {\MgII}, which is expected if the {\OVI} is collisionally
    ionized. In orientations in which outflows are expected, such as
    blue, face-on galaxies probed along the minor axis, the TPCFs are
    comparable between ions and show similar velocity dispersions.

  \item Given the vastly different kinematics of the {\OVI} absorbers
    compared to {\MgII}, it is very likely that the two ions trace
    different components of the CGM. In fact, the seven galaxies that
    have both detected {\OVI} and {\MgII} show values of $z_{abs}$
    (optical depth-weighted median of absorption) that can be offset
    by up to 100~{\kms}.

  \item The {\OVI} absorbers have similar velocity dispersions
    (Chi-squared test result of $<1\sigma$) regardless of galaxy
    color, inclination angle, and azimuthal angle, indicating that the
    gas is not strongly influenced by the present star formation
    activity in the galaxy (i.e., possibly deposited into the CGM by
    ``ancient outflows''). This is despite the findings of Kacprzak15
    that {\OVI} is preferentially observed along the projected major
    and minor axes of galaxies, which are frequently associated with
    accretion and outflows, respectively.

  \item The TPCF of the minor axis, edge-on subsample (likely
    dominated by outflowing material) may be more extended than the
    major axis, edge-on subsample ($4.6\sigma$). This large velocity
    separation tail is mostly due to a single large equivalent width
    absorber with a large velocity spread, which is classified as a
    large-scale galactic outflow by \citet{muzahid15}. Removing the
    outlier still results in a significant difference ($3.4\sigma$),
    however this significance is driven more by the major axis,
    edge-on subsample TPCF being slightly more narrow. Thus, the
    significant differences found in Figure~\ref{fig:PAincl} are
    likely not real.

  \item The uniform {\OVI} kinematics with galaxy color and
    orientation, and the azimuthal location preference of the {\OVI}
    absorbers point to ionization effects in the CGM. The gas giving
    rise to {\OVI} may be uniformly distributed throughout the CGM,
    but lower densities in intermediate azimuthal angles
    ($30^{\circ}<\Phi <60^{\circ}$) where outflows and accretion are
    less likely to occur may result in the oxygen being ionized out of
    the {\OVI} phase.

\end{enumerate}

To better understand how the kinematics of {\MgII} and {\OVI} compare,
it would be useful to form a large sample of galaxies with both {\OVI}
and {\MgII} absorption detected in high resolution quasar spectra. As
our samples stand now, it is not straightforward to compare the two
ion samples to each other due to differing galaxy property
distributions for the separate absorbing samples. Measuring the
multiphase kinematics for a set of galaxies like this is necessary for
understanding the multiphase CGM.

Lastly, observing the CGM in higher ionization states as a function of
galaxy properties is important if the azimuthal angle distribution of
{\OVI} is mainly due to differing ionization conditions. The higher
ionization states may also show an azimuthal preference, where the
incidence rate of higher ionization absorbers may be larger in the
intermediate azimuthal angles than {\OVI}. While it is not currently
feasible to study ions such as {\OVII} and {\OVIII} because they are
located in the x-ray regime, {\NeVIII} is observable in the UV. This
kinematics--galaxy orientation study could be done with {\NeVIII} once
a large enough sample is obtained.

%%%%%%%%%%%%%%%%%%%%%%%%%%%%%%%%%%%%
\acknowledgments

Support for this research was provided by NASA through grants HST
GO-13398 from the Space Telescope Science Institute, which is operated
by the Association of Universities for Research in Astronomy, Inc.,
under NASA contract NAS5-26555. G.G.K.~acknowledges the support of
the Australian Research Council through the award of a Future
Fellowship (FT140100933). M.T.M.~thanks the Australian Research
Council for Discovery Project grant DP130100568 which supported this
work.

\bibliographystyle{apj}
\bibliography{refs}

\end{document}